\documentclass[aps,twocolumn,showpacs,preprintnumbers,amsmath,amssymb,nofootinbib,superscriptaddress]{revtex4}

\usepackage{graphicx}
\usepackage{dcolumn}% Align table columns on decimal point
\usepackage{bm}% bold math%
\usepackage{amsmath}
\usepackage{hyperref}
\def\mk{\mathbf{k}}
\def\mq{\mathbf{q}}
\def\mA{\mathbf{A}}
\def\Sigmatau{\Sigma^{\tau_z}}

\def\mathE{\mathcal{E}}

\def\detbeta{\mathrm{det}(\widetilde{\beta})}

\def\Re{\mathrm{Re}}
\def\Im{\mathrm{Im}}

\newcommand{\secref}[1]{Sec.~\ref{#1}}
\newcommand{\eqnref}[1]{Eq.~(\ref{#1})}% or {(\ref{#1})}

\begin{document}

\title{
Topological-Berry-phase-induced spin torque current in a
two-dimensional system with generic $k$-linear spin-orbit
interaction}

\author{Tsung-Wei Chen}
\email{twchen@mail.nsysu.edu.tw}\affiliation{Department of
Physics, National Sun Yat-sen University, Kaohsiung 80424, Taiwan}

\author{Jian-Huang Li}\affiliation{Department of
Physics, National Sun Yat-sen University, Kaohsiung 80424, Taiwan}

\author{Chong-Der Hu}
\affiliation{Department of Physics and Center for Theoretical
Sciences, National Taiwan University, Taipei 106, Taiwan}

%\author{Dah-Wei Chiou}
%\email{dahwei@berkeley.edu}
%\affiliation{Department of Physics and Center for Theoretical
%Sciences, National Taiwan University, Taipei 106,
%Taiwan}\affiliation{Center for Condensed Matter Sciences, National
%Taiwan University, Taipei 106, Taiwan}

\date{\today}

\begin{abstract}
The Berry phase on the Fermi surface and its influence on the
conserved spin current in a two-dimensional system with generic
$k$-linear spin-orbit interaction are investigated. We calculate
the response of the effective conserved spin current to the
applied electric field, which is composed of conventional and spin
torque currents, by using the Kubo formula. We find that the
conventional spin current is not determined by the Berry phase
effect. Remarkably, the spin torque Hall current is found to be
proportional to the Berry phase, and the longitudinal spin torque
current vanishes because of the Berry phase effect. When the
$k$-linear spin-orbit interaction dominates the system, the Berry
phase on the Fermi surface maintains two invariant properties. One
is that the magnitude of the spin torque current protected by the
Berry phase is unchanged by a small fluctuation of energy
dispersion. The other one is that the change in the direction of
the applied electric field does not change the magnitude of the
spin torque current even if the energy dispersion is not
spherically symmetric; i.e., the Berry phase effect has no
dependence on the two-dimensional material orientation.  The spin
torque current is a universal value for all $k$-linear systems,
such as Rashba, Dresselhaus, and Rashba-Dresselhaus systems. The
topological number attributed to the Berry phase on the Fermi
surface represents the phase of the orbital chirality of spin in
the $k$-linear system. The change in the topological number
results in a phase transition in which the orbital chirality of
spin $s_z$ and $-s_z$ is exchanged. We found that the spin torque
current can be experimentally measured.

\end{abstract}
\maketitle

\section{Introduction}\label{sec:Intro}

The spin Hall effect, a phenomenon in which an applied electric
field generates a lateral spin current, has become an important
field in spintronics~\cite{pri98,Zutic04}. In the extrinsic spin
Hall effect, the transverse spin current occurs via spin-dependent
scattering of carriers with localized impurities via the
spin-orbit interaction~\cite{Hisch99}.  An intrinsic spin Hall
effect has also been proposed: Here, the applied longitudinal
electric field in an intrinsic spin-orbit-coupled system such as a
p-type or an n-type semiconductor  generates the transverse spin
current~\cite{Mura03,Sinova04}. Both intrinsic and extrinsic spin
Hall effects have been experimentally detected by various optical
measurements of spin accumulation~\cite{Kato04} and electrical
measurements through the inverse spin Hall effect~\cite{Saitoh06}.
The comparison of experimental results and theoretical
calculations has also been  thoroughly investigated by many
authors~\cite{theo}.

One of the potential applications of the spin Hall effect is to make
the spin current, if caused by the Berry curvature~\cite{Xiao10},
 dissipationless or independent of the impurity
scattering. The Berry phase stems from the accumulation by
adiabatic motion of electrons or holes on the Fermi surface, which
is the integration of the Berry curvature over the Brillouin zone.
If the Fermi level lies on the conduction band or valence bands,
the Berry curvature could provide a transverse Lorentz force in
momentum space such that the transverse spin current does not
cause Ohmic heat. When the bulk is gapped by the spin-orbit
interaction, the edge states caused by the Berry curvature can
carry spin current, which leads to the topological quantization of
the spin Hall effect~\cite{Xiao2006}.

In the Rashba and Rashba-Dresselhaus system~\cite{Rashba,Dre}, it
has been shown that the conventional definition of the spin
current, which is the symmetrical product of the velocity and
spin, is closely related to the Berry
phase~\cite{Shen04,Fujita10}. Nevertheless, as has been shown, a
small Dresselhaus spin-orbit interaction added to the Rashba
system leads to a nonvanishing longitudinal part of the
conventional spin conductivity~\cite{Sini04}. The influence of the
Berry phase on the bulk spin current depends not only on the
definition of the spin current but also on the position of the
Fermi level. Since the total spin is not conserved in a
spin-orbit-coupled system, the definition of a spin current is not
unique~\cite{DefSpin}. Moreover, it remains  unclear whether the
topological-Berry-phase-protected bulk spin current should be
unaltered by a small fluctuation of energy dispersion when the
Fermi level lies on the conduction or valence bands.

Recently, it has been shown that a nontrivial $\pi$ Berry phase
in the bulk Rashba semiconductor BiTeI can be detected by an
analysis of the Shubnikov--de Haas effect~\cite{Murakawa2013}. It
is important to investigate whether the distortion of energy
dispersion on the Fermi surface would break the Berry phase.

The topological number attributed to the Berry phase
plays an important role in the phase transition of the spin Hall
system. It has been show that, similar to the quantum Hall
effect~\cite{Tho82}, several plateau states exhibited by the spin
current in the kagome lattice can be ascribed to the Berry phase effect on
the Fermi surface~\cite{Liu09}. However, the mechanism of the
phase transition exhibited by the change in the topological number
is still unclear.

Motivated by these issues, in this paper, we focus on the
effective conserved bulk spin current proposed in
Ref.~\cite{Shi06}, which is composed of conventional spin current
$J^{s_z}_a=\frac{1}{2}\left\{s_z,v_a\right\}$ and spin torque
current $J^{\tau_z}_a=\frac{1}{2}\left\{ds_z/dt,x_a\right\}$.  The
total spin current $J^{z}_a=J^{s_z}_a+J^{\tau_z}_a$ effectively
satisfies the continuity equation
\begin{equation}
\frac{\partial\mathcal{S}_z}{\partial
t}+\nabla_a\mathcal{J}^z_a=0,
\end{equation}
where $\mathcal{S}_z=\Psi^{\dag}s_z\Psi$ is the spin density and
$\mathcal{J}^z_a=\Re\left(\Psi^{\dag}J^z_a\Psi\right)$ is the spin
current density. The wave function $\Psi$ satisfies the
Schr\"odinger equation $H\Psi=i\hbar\partial\Psi/\partial t$,
where $H=H_0+e\mathbf{E}\cdot\mathbf{x}$, with $\mathbf{E}$ being
the applied electric field. We also focus on the two-dimensional
spin-orbit-coupled system with a Fermi level lying on the
conduction or valence band. The Berry phase is obtained on the
Fermi surface. We study the generic $k$-linear spin-orbit-coupled
system in which energy dispersion could be nonspherical and the
fluctuation of energy dispersion can be studied by changing the
spin-orbit strength. We find that the Berry-phase-protected
response is invariant under a fluctuation of energy dispersion. A
small fluctuation of the energy dispersion changes the magnitude
of the conventional spin Hall response and breaks the
antisymmetric properties of the conventional spin conductivity.
The conventional spin conductivity is closely related to the Berry
phase but not protected by the Berry phase effect. Remarkably, we
find that the spin torque current is protected by the Berry phase
effect. Moreover, we find that the topological number induced by
the Berry phase on the Fermi surface manifests the phase
transition of the orbital chirality of spin.

Our present paper is organized as follows. In
\secref{sec:SpinDynamics}, to simplify the calculation, we rewrite
the Kubo formulas of conventional and spin torque conductivity
tensors in terms of the spin-orbit interaction of the
two-dimensional system. In \secref{sec:Berryphase}, we study the
generic $k$-linear system and calculate the Berry phase on the
Fermi surface. We show that the Berry phase on the Fermi surface
manifests two invariant properties. We also show that the
topological number induced by the Berry phase governs the orbital
chirality phase transition of the spin current. In
\secref{sec:Berry-Spin}, we calculate the conventional and spin
torque conductivities. We find that the conventional spin current
is not protected by the Berry phase. The spin torque conductivity
is shown to be proportional to the Berry phase and the
longitudinal spin torque conductivities vanish. Furthermore, the
spin torque Hall conductivity satisfies antisymmetric properties.
The effect of disorder is also addressed. Our conclusions are
presented in \secref{sec:Conclusion}. Some calculations are
provided in the Appendices.

\section{conserved spin conductivity tensor}\label{sec:SpinDynamics}
In this section, we simply review the Kubo formulae of the
conventional spin current and spin torque current in the
two-dimensional spin-orbit-coupled system. The system Hamiltonian
in the presence of an applied electric field is given by
\begin{equation}\label{TotalH}
H=H_0+e\mathbf{E}\cdot\mathbf{x},
\end{equation}
where the external perturbation is the in-plane electric field
$e\mathbf{E}\cdot\mathbf{x}$, where $e>0$ and $-e$ is the electric
charge of an electron. The unperturbed Hamiltonian $H_0$ is the
two-dimensional spin-orbit-coupled system,
\begin{equation}\label{H0}
H_0=\varepsilon_{\mk}+\sigma_xd_x+\sigma_yd_y,
\end{equation}
where $\varepsilon_{\mk}=\hbar^2k^2/2m$, $\sigma_i$ ($i=x,y,z$)
are Pauli matrices, and $d_i$ ($i=x,y$) are functions of $\mk$ and
the spin-orbit interaction. The eigenvalue equation of Eq. (\ref{H0}),
$H_0|n\mk\rangle=\mathcal{E}_{n\mk}|n\mk\rangle,$ can be solved
exactly. The eigenvalue is given by
\begin{equation}\label{eigenenergy}
\mathcal{E}_{n\mk}=\varepsilon_{\mk}-nd
\end{equation}
with $d=\sqrt{d_x^2+d_y^2}$ and $n=\pm1$. The eigenstate of the
Hamiltonian  (\ref{H0}) can be chosen as
\begin{equation}\label{eigenvector}
|n\mk\rangle=\frac{1}{\sqrt{2}}\left(\begin{array}{c}
e^{-i\theta}\\
in
\end{array}\right),
\end{equation}
where $\theta=\tan^{-1}(-d_y/d_x)$ is also a function of $\mk$. It
can be shown that $\langle n\mk|\sigma_z|n\mk\rangle=0$ and
$\langle n\mk|\boldsymbol{\sigma}|n\mk\rangle=-n\mathbf{d}/d$. In
the absence of an electric field, the direction of spin on the
Fermi surface is along the vector
$\mathbf{d}=d_x\hat{e}_x+d_y\hat{e}_y$.

The total spin current linear response to the electric field is
given by
\begin{equation}
\left(\begin{array}{c} J^z_x\\
J^z_y
\end{array}\right)=\left(\begin{array}{cc}
\sigma_{xx}^z&\sigma_{xy}^z\\
\sigma_{yx}^z&\sigma_{yy}^z
\end{array}\right)\left(\begin{array}{c}
E_x\\
E_y
\end{array}\right),
\end{equation}
where the total spin conductivity tensor $\sigma^z_{ab}$ is
composed of the conventional spin conductivity tensor
$\sigma^{s_z}_{ab}$ and the spin torque conductivity tensor
$\sigma^{\tau_z}_{ab}$:
\begin{equation}
\sigma^z_{ab}=\sigma^{s_z}_{ab}+\sigma^{\tau_z}_{ab}.
\end{equation}
The Kobo formula for the conventional spin conductivity tensor is
given by
\begin{equation}\label{Kubo-ConSpin}
\begin{split}
\sigma_{ab}^{s_z}=&\frac{e\hbar}{V}\sum_{n\neq
n'}\sum_{\mk}\frac{f_{n\mk}-f_{n'\mk}}{(\mathE_{n\mk}-\mathE_{n'\mk})^2}\\
&\times\Im\langle n\mk|J_a^{s_z}|n'\mk\rangle\langle
n'\mk|v_b|n\mk\rangle.
\end{split}
\end{equation}
The Kubo formula for the spin torque conductivity tensor is given
by~\cite{k3hole}
\begin{equation}\label{Kubo-Torspin-1}
\begin{split}
\sigma^{\tau_z}_{ab}=&\lim_{q_a\rightarrow0}\frac{1}{q_a}\frac{e\hbar}{V}\sum_{n\neq
n'}\sum_{\mk}\frac{f_{n\mk}-f_{n'\mk+\mq}}{(\mathE_{n\mk}-\mathE_{n'\mk+\mq})^2}\\
&\times\Re\langle n\mk|\tau_z(\mk,\mq)|n'\mk+\mq\rangle\langle
n'\mk+\mq|v_b(\mk,\mq)|n\mk\rangle,
\end{split}
\end{equation}
where $\tau_z(\mk,\mq)=[\tau_z(\mk)+\tau_z(\mk+\mq)]/2$ and
$v_b(\mk,\mq)=[v_b(\mk)+v_b(\mk+\mq)]/2$ with torque
$\tau_z(\mk)=(1/i\hbar)[s_z,H_0]$ and velocity $v_b(\mk)=\partial
H_0/\hbar\partial k_b$.

It can be shown that the conventional spin conductivity [Eq.
(\ref{Kubo-ConSpin})] can be written in terms of $d_a$
\cite{Bern05} as
\begin{equation}\label{ConSpin}
\sigma^{s_z}_{ab}=\frac{e}{V}\sum_{\mk}(f_{\mk+}-f_{\mk-})\left(\frac{\partial\varepsilon_{\mk}}{\partial
k_a}\right)\frac{1}{4d^3}\left(d_x\frac{\partial d_y}{\partial
k_b}-d_y\frac{\partial d_x}{\partial k_b}\right).
\end{equation}
It has been shown that the polarized spin response,
\begin{equation}\label{mathQ}
\mathcal{Q}^z_{a}=\frac{e\hbar}{4d^3}\left(d_x\frac{\partial
d_y}{\partial k_a}-d_y\frac{\partial d_x}{\partial k_a}\right),
\end{equation}
is equivalent to the magnitude of the spin projecting on the
out-of-plane magnetic field,
$\mathcal{Q}^z_a\equiv(\hbar/2)B_z/|\mathbf{B}|,$ with
$|\mathbf{B}|=d$ and
$B_z=e(\mathbf{d}\times\partial\mathbf{d}/\partial k_a)_z/2d^2$,
which can reproduce the result of Refs.~\cite{Sinova04,Fujita10}
in the Rashba system.

The spin torque conductivity tensor [Eq. (\ref{Kubo-Torspin-1})]
can also be written in terms of $d_a$ and simplified to the
following form (see Appendix~\ref{App-TorSpin}):
\begin{equation}\label{TorSpin}
\sigma^{\tau_z}_{ab}=-2\sigma^{s_z}_{ab}+\sigma^{s_z}_{ba}+\Sigmatau_{ab},
\end{equation}
where the pure spin torque conductivity $\Sigmatau_{ab}$ is given
by
\begin{equation}\label{Sigmatau}
\Sigmatau_{ab}=-\frac{e}{2V}\sum_{n\mk}\frac{\partial
f_{n\mk}}{\partial k_a}\frac{1}{2d^2}\left(d_x\frac{\partial
d_y}{\partial k_b}-d_y\frac{\partial d_x}{\partial k_b}\right).
\end{equation}
Equation (\ref{TorSpin}) has also been found in Ref.~\cite{Liu09}
in the two-band model with Rashba spin-orbit interaction. In fact,
from the Kubo formula  (\ref{Kubo-Torspin-1}), it can be shown
that the response $\Sigmatau_{ab}$ is obtained from the following
Kubo formula near the Fermi surface:
\begin{equation}
\Sigmatau_{ab}=-\frac{e\hbar}{V}\sum_{n\neq
n'}\sum_{\mk}\frac{\partial f_{n\mk}}{\partial
k_a}\frac{\Re\langle n\mk|\tau_z|n'\mk\rangle\langle
n'\mk|v_b|n\mk\rangle}{(\mathE_{n\mk}-\mathE_{n'\mk})^2},
\end{equation}
which is the electric-field-induced pure spin torque $\tau_z$ on
the Fermi surface. We note that the spin torque conductivity
tensor is in general not twice as large as the conventional spin
conductivity tensor and opposite in sign \cite{TWChen09}. In the
following sections, we first consider a generic $k$-linear
spin-orbit-coupled system, in which the energy dispersion is
nonspherical. Furthermore, we investigate the relationship between
the Berry phase and the spin current by using Eqs. (\ref{ConSpin})
and (\ref{TorSpin}).

\section{Berry Phase in a $k$-linear spin-orbit-coupled system}\label{sec:Berryphase}
\subsection{Generic $k$-linear system}
Consider a generic $k$-linear spin-orbit-coupled system, where
$d_x=\beta_{xx}k_x+\beta_{xy}k_y$,
$d_y=\beta_{yx}k_x+\beta_{yy}k_y$, and $d=k\Gamma(\phi)$ with
\begin{equation}
\begin{split}
\Gamma(\phi)^2=&(\beta_{xx}^2+\beta_{yx}^2)\cos^2\phi+(\beta_{xy}^2+\beta_{yy}^2)\sin^2\phi\\
&+(\beta_{xx}\beta_{xy}+\beta_{yx}\beta_{yy})\sin(2\phi).
\end{split}
\end{equation}
The spin-orbit interaction term $\sigma_xd_x+\sigma_yd_y$ can be
written as $\sum_{ij}\sigma_i\beta_{ij}k_j$, where the spin-orbit
matrix $\beta_{ij}$ represents the spin-orbit interactions in the
system. Some mechanisms could result in $k$-linear spin-orbit
interaction. Structure inversion asymmetry (SIA) results in
a pure Rashba spin-orbit interaction~\cite{Rashba}, and the
spin-orbit matrix elements are $\beta_{xx}=\beta_{yy}=0$ and
$\beta_{xy}=-\beta_{yx}=\alpha$. Bulk inversion asymmetry
(BIA) results in  pure Dresselhaus spin-orbit
interactions~\cite{Dre}, and the spin-orbit matrix elements are
$\beta_{xx}=-\beta_{yy}=\gamma$ and $\beta_{xy}=\beta_{yx}=0$. The
strain effect in SIA and BIA systems can also induce  $k$-linear
spin splitting that is linear in momentum~\cite{BernePRB05}.

\begin{figure}
\begin{center}
\includegraphics[width=8cm,height=4cm]{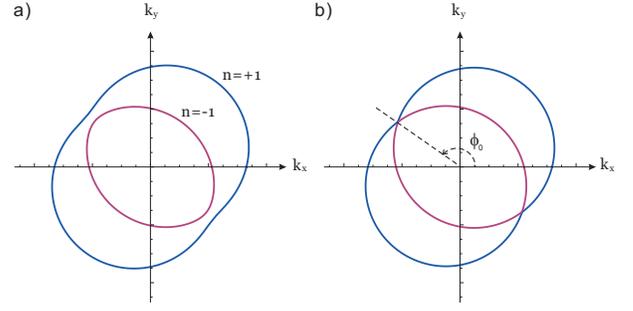}
\end{center}
\caption{(Color online) Energy dispersion of a $k$-linear system
on the Fermi surface showing (a) $\detbeta\neq0$ and (b)
$\detbeta=0$, for which two degenerate points appear at $\phi_0$
and $\pi-\phi_0$.}\label{FigEngDis}
\end{figure}

The terms $d_x(\partial d_y/\partial k_b)-d_y(\partial
d_x/\partial k_b)$ appearing in the conventional and spin torque
conductivities in the $x$ and $y$ components are
\begin{equation}\label{dpd}
\begin{split}
&d_x\frac{\partial d_y}{\partial k_x}-d_y\frac{\partial
d_x}{\partial k_x}=-\detbeta k_y,\\
&d_x\frac{\partial d_y}{\partial k_y}-d_y\frac{\partial
d_x}{\partial k_y}=\detbeta k_x,
\end{split}
\end{equation}
where
\begin{equation}\label{detbeta}
\detbeta=\det\left(\begin{array}{cc}
\beta_{xx}&\beta_{xy}\\
\beta_{yx}&\beta_{yy}
\end{array}\right)=\beta_{xx}\beta_{yy}-\beta_{yx}\beta_{xy}.
\end{equation}
Both the conventional spin current and the spin torque current
contain Eq. (\ref{dpd}) in the integral. When $\detbeta=0$, both
conventional spin and spin torque currents should vanish. This can
be seen as follows. When degeneracy occurs by tuning the
spin-orbit strength such that $\Gamma(\phi_0)=0$ at some angle
$\phi_0$, as shown in Fig. \ref{FigEngDis}~\cite{para}, we find
that the angle $\phi_0$ is determined by
$\tan(\phi_0)=-[(\beta_{xx}\beta_{xy}+\beta_{yx}\beta_{yy})\pm\sqrt{-|\detbeta|^2}]/(\beta_{xy}^2+\beta_{yy}^2)$.
Therefore, the occurrence of degeneracy that leads to the result
$\detbeta=0$ implies that the system should have an additional
conserved quantity. We find that the conserved quantity (a unitary
matrix), up to a global phase, is given by
\begin{equation}\label{Uspin}
\mathcal{U}=\sqrt{\frac{1+A}{2}}\sigma_x+\sqrt{\frac{1-A}{2}}\frac{B}{|B|}\sigma_y,
\end{equation}
where
$A=(\beta_{xy}\beta_{xx}-\beta_{yx}\beta_{yy})/(\beta_{xy}\beta_{xx}+\beta_{yx}\beta_{yy})$
and
$B=(\beta_{xx}\beta_{yy})/(\beta_{xy}\beta_{xx}+\beta_{yx}\beta_{yy})$.
It can be shown that $|B|=\sqrt{1-A^2}$. For example, in the
Rashba-Dresselhaus system, the spin-orbit matrix elements are
$\beta_{xx}=\beta$, $\beta_{xy}=\alpha$, $\beta_{yx}=-\alpha$, and
$\beta_{yy}=-\beta,$ and we have $A=0$ and $B=-\beta/\alpha$. For
$\alpha=\beta$, we have
$\mathcal{U}=(\sigma_x-\sigma_y)/\sqrt{2}$, and, for
$\alpha=-\beta$, we have
$\mathcal{U}=(\sigma_x+\sigma_y)/\sqrt{2}$. The system Hamiltonian
with $\detbeta=0$ is invariant under the unitary transformation,
i.e., $\mathcal{U}H_0\mathcal{U}^{\dag}=H_0$. We find that the
in-plane spin ($\sigma_x,\sigma_y$) under the unitary
transformation [\eqnref{Uspin}] can be given by
\begin{equation}
\begin{split}
&\mathcal{U}\sigma_x\mathcal{U}^{\dag}=A\sigma_x+B\sigma_y,\\
&\mathcal{U}\sigma_y\mathcal{U}^{\dag}=B\sigma_x-A\sigma_y.
\end{split}
\end{equation}
For the $z$ component, interestingly, we find that under the unitary
transformation $\sigma_z$ is simply replaced by $-\sigma_z$, i.e.,
\begin{equation}
\mathcal{U}\sigma_z\mathcal{U}^{\dag}=-\sigma_z.
\end{equation}
This implies that the spin current $J^z_a$ (transverse and
longitudinal) in the original basis has the same magnitude as
that in the transformed basis but they are opposite in sign.
Nevertheless, the system with $\detbeta=0$ has the same
Hamiltonian in both basis, and, thus, this leads to the result that
the spin current $J_a^z$ must vanish when $\detbeta=0$.

The vanishing spin current can also be seen as follows. It has
been shown that, in the generic $k$-linear system, the effective
coupling between spin $z$ component $s_z$ and orbital angular
momentum $L_z$ is given by $(-2m/\hbar^4)\detbeta
s_zL_z$~\cite{Valin11,TWChen13}. When $\detbeta=0$, the effective
coupling between spin and orbital angular momentums in the $z$
component vanishes;  thus, both conventional and spin torque
currents must be zero. The effective coupling
$(-2m/\hbar^4)\detbeta$ will play an important role in the
interpretation of phase transition caused by the Berry phase, as
shown in the following section.

\subsection{Berry phase on the Fermi surface}
By using Eq. (\ref{eigenvector}), we find that the Berry vector
potential $A_a(\mk)$ can be written as\footnote{If we use
a different basis, the Berry vector potential could depend on the
band index. However, it does not change the physics conclusion in
this paper since the integrant in Eq. (\ref{intA}) is invariably
unchanged.}
\begin{equation}\label{BerryA}
\begin{split}
A_a&=\langle n\mathbf{k}|i\frac{\partial}{\partial k_a}|n\mk\rangle\\
&=\frac{1}{2}\frac{\partial\theta}{\partial k_a}\\
&=\frac{1}{2d^2}\left(d_x\frac{\partial d_y}{\partial
k_a}-d_y\frac{\partial d_x}{\partial k_a}\right).
\end{split}
\end{equation}
Equation (\ref{dpd}) also implies that the Berry vector potential
[Eq. (\ref{BerryA})] is given by
\begin{equation}\label{BerryA-linear}
\mathbf{A}=\frac{1}{2}\frac{\partial\theta}{\partial\mk}=\frac{\detbeta(-k_y\hat{e}_x+k_x\hat{e}_y)}{2k^2\Gamma(\phi)^2}.
\end{equation}
However, in spherical
coordinates, the Berry vector potential   can be written as
$\mathbf{A}=A_{\rho}\hat{e}_{\rho}+A_{\phi}\hat{e}_{\phi}$ with
$A_{\rho}=\mk\cdot\mA/k=0$ and
$A_{\phi}=(\mk\times\mA)_z/k=\detbeta/2k\Gamma(\phi)$. Therefore,
in a generic Dirac Hamiltonian the Berry vector potential has only an
$A_{\phi}$ component. By using Eq. (\ref{BerryA-linear}) and the line
element
$d\boldsymbol{\ell}=dk\hat{e}_{\rho}+kd\phi\hat{e}_{\phi}$, the
Berry phase $\Phi$ is given by
\begin{equation}\label{intA}
\begin{split}
\Phi&=\oint \mathbf{A}\cdot
d\boldsymbol{\ell}\\
&=\frac{1}{2}\detbeta\int_0^{2\pi}\frac{d\phi}{\Gamma(\phi)^2}=\pi\frac{\detbeta}{|\detbeta|}.
\end{split}
\end{equation}
We note that the integral is performed on the Fermi surface. In
obtaining Eq. (\ref{intA}), we have used the result
$\int_0^{2\pi}d\phi(1/\Gamma(\phi)^2)=2\pi/|\detbeta|$, which can
be obtained by using the residue method to the two poles
$\lambda_1=(\beta_{xx}+\beta_{yy})+i(\beta_{yx}-\beta_{xy})$ and
$\lambda_2=(\beta_{xx}-\beta_{yy})+i(\beta_{yx}+\beta_{xy})$. The
Berry curvature $F^z_{ab}$ is defined as
\begin{equation}\label{BerryB}
F^{z}_{ab}=\frac{\partial A_b}{\partial k_b}-\frac{\partial
A_a}{\partial k_b}.
\end{equation}
For $\mk\neq0$, we have $\nabla_{\mk}\times\mA=0$, and the Berry
curvature vanishes everywhere except at $\mk=0$. Using the divergence
theorem in two dimensions and taking the small closed curve around
the point $\mk=0$, we obtain
\begin{equation}\label{BerryB-linear}
F^z_{xy}=-F^z_{yx}=\Phi\delta(\mk).
\end{equation}
Since the Berry curvature in the generic $k$-linear system is a
delta function peaked at $\mk=0$, when $\mk\neq0$, the Berry
curvature vanishes. As a result, the transverse current of spin
cannot be caused by the Lorentz force in  momentum space. However,
the spin torque current by definition does not require the charge
current in which $\mk=0$ states would have a large contribution to
the spin torque current, as will be shown in the following
section.

The Berry phase $\Phi$ is an invariant quantity in the sense that
$\detbeta$ is invariant under rotation. That is, if the response
is caused by the Berry phase, the response remains a universal
constant regardless of the change in the direction of applied
electric field. Furthermore, the Berry phase $\Phi$ is a constant
$\pm\pi$, which is independent of spin-orbit strength. This means
that a small change in the spin-orbit strength $\beta_{ij}$ does
not change the value of $\Phi$ and so the Berry-phase-induced response does not change its magnitude under a
fluctuation of energy dispersion.

\begin{figure}
\begin{center}
\includegraphics[width=6cm,height=4cm]{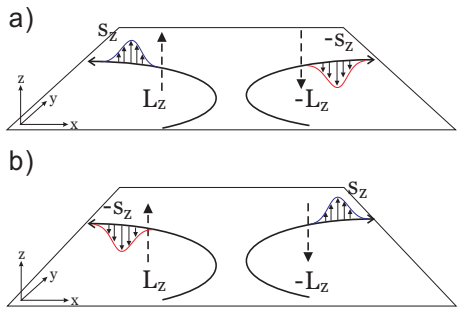}
\end{center}
\caption{(Color online) Schematic diagram showing the phase
transition in a two-dimensional spin Hall system from (a)
$\detbeta>0$ to (b) $\detbeta<0$.}\label{FigDetB}
\end{figure}

In the quantum Hall effect~\cite{Tho82}, the topological number
exhibited by the Berry phase divided by $2\pi$ is an integer.
Similar to the quantum Hall effect, the Berry
phase divided by $\pi$ is also an integer:
$\detbeta/|\detbeta|=\pm1$. The physical meaning of the integer in
the quantum Hall effect is the number of edge states, in which
the change in the topological number results in a phase
transition of a quantum Hall system from $n$ to $n\pm1$ edge
states. The physical meaning of the topological number
$\detbeta/|\detbeta|=\pm1$ is as follows. It has been shown that
 $\detbeta$ governs the effective coupling\footnote{The
coupling should be $-(2m/\hbar^4)\detbeta s_zL_z$. We neglect the
overall minus sign for convenience of the discussion.} of $s_z$
and $L_z$ via $\detbeta s_zL_z$~\cite{Valin11, TWChen13} and,
thus, represents the orbital chirality of the spin $s_z$. When
$\detbeta/|\detbeta|$ changes from $+1$ to $-1$, the current
carrying $+s_z$ changes orbital chirality from $L_z$ to $-L_z$,
and $-s_z$ changes orbital chirality from $-L_z$ to $+L_z$. The
phase transition is summarized in Fig.~\ref{FigDetB}. When
$\detbeta=0$, the spin Hall effect vanishes, as mentioned above.
Therefore, the change in the topological number of
$\detbeta/|\detbeta|$ results in the phase transition of
exchanging the orbital chirality of spin $s_z$ and $-s_z$ in the
spin Hall system.

In the following section, we will show that the conventional spin
conductivity is not protected by the Berry phase and that the spin
torque conductivity is truly due to the Berry phase effect.

\section{Berry phase and Spin current}\label{sec:Berry-Spin}

By using Eqs. (\ref{ConSpin}) and (\ref{dpd}), we obtain the four
components of the conventional spin conductivity tensor:
\begin{equation}\label{ConSpin-linear}
\begin{split}
&\sigma^{s_z}_{xy}=\frac{e}{8\pi^2}\detbeta\int_0^{2\pi}\frac{\cos^2\phi}{\Gamma(\phi)^2}d\phi,\\
&\sigma^{s_z}_{yx}=-\frac{e}{8\pi^2}\detbeta\int_0^{2\pi}\frac{\sin^2\phi}{\Gamma(\phi)^2}d\phi,\\
&\sigma^{s_z}_{xx}=-\sigma^{s_z}_{yy}=-\frac{e}{8\pi^2}\detbeta\int_0^{2\pi}\frac{\sin\phi\cos\phi}{\Gamma(\phi)^2}d\phi.
\end{split}
\end{equation}
Comparing Eq. (\ref{ConSpin-linear}) with Eq. (\ref{intA}), we
find that the conventional spin conductivity is not determined by
the Berry phase. However, it is interesting to evaluate each
integral in Eq. (\ref{ConSpin-linear}). The integral can be
performed by using the residue method to the two poles
$\lambda_1=(\beta_{xx}+\beta_{yy})+i(\beta_{yx}-\beta_{xy})$ and
$\lambda_2=(\beta_{xx}-\beta_{yy})+i(\beta_{yx}+\beta_{xy})$, and
the result is given by
%\begin{widetext}
\begin{equation}\label{ConSpinMatrix}
\begin{split}
&\left(\begin{array}{cc} \sigma^{s_z}_{xx}&\sigma^{s_z}_{xy}\\
\sigma^{s_z}_{yx}&\sigma^{s_z}_{yy}
\end{array}\right)\\
&=\frac{\detbeta}{|\detbeta|}\frac{e}{8\pi}\left(\begin{array}{cc}
\Im\left(\frac{\lambda_{<}}{\lambda_{>}}\right)&\left[1-\Re\left(\frac{\lambda_{<}}{\lambda_{>}}\right)\right]\\
-\left[1+\Re\left(\frac{\lambda_{<}}{\lambda_{>}}\right)\right]&-\Im\left(\frac{\lambda_{<}}{\lambda_{>}}\right)
\end{array}\right),
\end{split}
\end{equation}
%\end{widetext}
where $\lambda_{>}$ ($\lambda_{<}$) is taken from the relative
maximum (minimum) value of $(|\lambda_1|,|\lambda_2|)$. That is,
if $|\lambda_1|>|\lambda_2|$ then $\lambda_{>}=\lambda_1$ and
$\lambda_<=\lambda_2$, and vice versa~\cite{TWChen13}. The result
shows that the conventional spin conductivity seems to be affected
by the Berry phase. Because $\Re(\lambda_</\lambda_>)$ and
$\Im(\lambda_</\lambda_>)$ in Eq. (\ref{ConSpinMatrix}) depend on
the spin-orbit strength $\beta_{ij}$, we find that
$\sigma_{ab}^{s_z}$ is not purely caused by the Berry phase. The
Berry-phase-induced spin current should have vanishing
longitudinal conductivity.

In the pure Rashba system, we have $\Im(\lambda_</\lambda_>)=0$
and $\Re(\lambda_</\lambda_>)=0$. It seems that the conventional
spin conductivity is protected by the Berry phase. Nevertheless, a
small Dresselhaus spin-orbit interaction breaks the spherically
symmetric energy dispersion and leads to
$\Im(\lambda_</\lambda_>)\neq0$, which depends on the ratio
$\beta/\alpha$ or $\alpha/\beta$~\cite{Sini04}; i.e., the
longitudinal term is not zero.

In the Rashba-Dresselhaus system
$H_0=\varepsilon_{\mk}+\alpha(\sigma_xk_y-\sigma_yk_x)+\gamma(\sigma_xk_x-\sigma_yk_y)$,
the determinant of the spin orbit matrix is $\alpha^2-\gamma^2$,
and $\lambda_1=-2i\alpha$ and $\lambda_2=2\gamma$ (if the electric
field is applied in the direction $[010]$).  We have
$\Re(\lambda_</\lambda_>)=0$; however,
$\Im(\lambda_</\lambda_>)\neq0$. Although the conventional
spin Hall conductivity in the Rashba-Dresselhaus system is
numerically equal to the Berry phase (i.e., the system has nonzero
longitudinal terms), a small change in the direction of the applied
electric field actually results in
$\Re(\lambda_</\lambda_>)\neq0$. For example, when the electric
field is applied in $[1\bar{1}0]$, we change the coordinate
$(k_x,k_y)$ to $(k_x',k_y')$ such that $k_x'$ and $k_y'$ are
parallel to $[110]$ and $[1\bar{1}0]$, respectively. The resulting
spin-orbit matrix elements are
$\beta_{x'x'}=(\alpha-\gamma)/\sqrt{2}$,
$\beta_{x'y'}=(\alpha+\gamma)/\sqrt{2}$,
$\beta_{y'x'}=-(\alpha-\gamma)/\sqrt{2}$, and
$\beta_{y'y'}=(\alpha+\gamma)/\sqrt{2}$. It can be shown that the
determinant of the new spin-orbit matrix is still
$\alpha^2-\gamma^2$. We have $\lambda_1=\sqrt{2}\alpha(1+i)$ and
$\lambda_2=-\sqrt{2}\gamma(1-i)$, and
$\Re(\lambda_</\lambda_>)\neq0$. Therefore, the conventional spin
current in the Rashba-Dresselhaus system does not behave as an
isotropic system.

If the conventional spin current was protected by the Berry phase,
then the resulting conventional spin current should have
maintained a universal value under a small fluctuation of energy
dispersion (without breaking time-reversal symmetry). However, as
we have shown above, the conventional spin current depends on the
shape of the energy dispersion. The conventional spin current is
not purely caused by the Berry phase.

We now consider the spin torque conductivity tensor. We first
calculate $\Sigmatau_{ab}$ by using Eq. (\ref{Sigmatau}). It can
be shown that (see Appendix \ref{App-Sigmatau})
\begin{equation}\label{Sigmatau-linear}
\begin{split}
&\Sigmatau_{xx}=-\Sigmatau_{yy}=-\frac{e^2}{8\pi^2}\detbeta\int_0^{2\pi}\frac{\sin\phi\cos\phi}{\Gamma(\phi)^2}d\phi,\\
&\Sigmatau_{xy}=+\frac{e^2}{8\pi^2}\detbeta\int_0^{2\pi}\frac{\cos^2\phi}{\Gamma(\phi)^2}d\phi,\\
&\Sigmatau_{yx}=-\frac{e^2}{8\pi^2}\detbeta\int_0^{2\pi}\frac{\sin^2\phi}{\Gamma(\phi)^2}d\phi.
\end{split}
\end{equation}
Comparing Eq. (\ref{Sigmatau-linear}) with Eq.
(\ref{ConSpin-linear}), we find that
$\sigma^{s_z}_{ab}=\Sigmatau_{ab}$. Substituting Eqs.
(\ref{ConSpin-linear}) and (\ref{Sigmatau-linear}) into
(\ref{TorSpin}) and performing some straightforward calculations, we have
\begin{equation}\label{TorSpin-linear}
\begin{split}
&\sigma^{\tau_z}_{xy}=-\sigma^{\tau_z}_{yx}=-\frac{e}{8\pi^2}\detbeta\int_0^{2\pi}\frac{1}{\Gamma(\phi)^2}d\phi,\\
&\sigma^{\tau_z}_{xx}=\sigma^{\tau_z}_{yy}=0.
\end{split}
\end{equation}
Comparing Eq. (\ref{TorSpin-linear}) with the Berry phase given by
Eq. (\ref{intA}), we see that the spin torque Hall conductivity is
closely related to the Berry phase effect. By using Eq.
(\ref{intA}) we obtain
\begin{equation}\label{TorSpinMatrix}
\left(\begin{array}{cc} \sigma^{\tau_z}_{xx}&\sigma^{\tau_z}_{xy}\\
\sigma^{\tau_z}_{yx}&\sigma^{\tau_z}_{yy}
\end{array}\right)=\frac{e}{4\pi^2}\Phi\left(\begin{array}{cc}
0&-1\\
1&0
\end{array}\right).
\end{equation}
The magnitude of the spin torque Hall conductivity is independent
of the direction of the applied electric field. The spin torque
response of the system behaves like an isotropic system. That is,
a small change in the direction of the applied electric field does
not change the spin torque Hall response. In addition, a small
fluctuation of energy dispersion does not change the magnitude of
the spin torque current. Furthermore, the longitudinal
conductivities are zero.  By using the Berry curvature [Eq.
(\ref{BerryB-linear})] and
$\sigma^{\tau_z}_{ab}=-\frac{e}{2V}\sum_{n\mk}f_{n\mk}F^z_{ab}$
with $\mu>0$, we can also reproduce the result of Eq.
(\ref{TorSpinMatrix}). This also explains why the longitudinal
spin torque conductivities vanish and the transverse part
satisfies the antisymmetric property.

%\begin{figure}
%\begin{center}
%\includegraphics[width=6.5cm,height=5.5cm]{FigExp.eps}
%\end{center}
%\caption{Schematic diagram showing the spin-Hall effect. The wave
%packet envelop with arrows represents torque-spin current. The
%filled circle with an arrow stands for the conventional spin. a)
%the total spin current generated by electric current. b) only the
%spin torque current generated by electric field.}\label{FigExp}
%\end{figure}

However, because the Fermi level does not lie in the true gap, the
conserved spin current may be affected by  impurity scattering.
The conventional spin conductivity in the Rashba system is shown
to be significantly influenced by  impurity
scattering~\cite{Loss05}. Impurity scattering in the Rashba system
with the  spin torque current taken into account has been studied
by Sugimoto \textit{et al}.~\cite{Sugi06}. By using the Keldysh
formalism, the conserved spin Hall current is shown to be zero in
the Rashba system with a delta impurity potential and remains a
finite value with a finite range potential.

In short, the spin torque current protected by the Berry phase is
unaltered by a fluctuation of energy dispersion and a change in
the direction of the applied electric field. The
Berry-phase-induced spin torque current does not contribute to
dissipation. Unlike the quantum Hall effect, the topological
number may not be invariant under the influence of impurity
scattering.

We propose an experiment for detecting our perdition of
rotationally invariant spin torque current. By using Eqs.
(\ref{ConSpinMatrix}) and (\ref{TorSpinMatrix}), the total
spin-Hall conductivity
$\sigma^z_{xy}=\sigma^{s_z}_{xy}+\sigma^{\tau_z}_{xy}$ satisfies
the following equation
\begin{equation}
\frac{\sigma^z_{xy}+\sigma^z_{yx}}{\sigma^z_{xy}-\sigma^z_{yx}}=\lambda,
\end{equation}
where $\lambda=\Re(\lambda_</\lambda_>)$.  The experimental value
of the spin current is assumed to be the sum of conventional and
spin torque currents, and we could obtain the corresponding
experimental value $\lambda_{\exp}$. The theoretical value of the
conventional spin-Hall conductivity should be
$\sigma^{s_z}_{xy,\mathrm{th}}=\pm(e/8\pi)(1-\lambda_{\exp})$. We
can calculate the theoretical value of the spin torque-Hall
conductivity
\begin{equation}
\sigma^{\tau_z}_{xy,\mathrm{th}}=\sigma^z_{xy,\mathrm{exp}}-\left(\pm\frac{e}{8\pi}\right)\left(1-\lambda_{\exp}\right).
\end{equation}
If we rotate the direction of an applied electric field, we should
obtain different values of $\lambda_{\exp}$ and
$\sigma^z_{xy,\mathrm{exp}}$ in a linear momentum dominate regime
with nonspherical symmetric energy dispersion, such as the Rashba
system with a small correction of Dresselhaus spin-orbit strength
in II-VI semiconductors~\cite{Rashba, Zutic04}. However, the
resulting $\sigma^{\tau_z}_{xy,\mathrm{th}}$ should be
rotationally invariant. We predict that
$\sigma^{\tau_z}_{xy,\mathrm{th}}$ maintains a universal constant
and is experimentally measurable.

So far we have not addressed the contribution of higher momentum
to the Berry phase and spin torque current. The Berry curvature in
a two-dimensional system is always a Dirac delta function at
$\mk=0$ since the Berry vector potential is a gradient of the
scalar function $\theta(\mk)$ [see Eq. (\ref{BerryA})]. For the
$k$-cubic Rashba system~\cite{Winkler02}, we have
$d_x=\alpha_hk^3\sin(3\phi)$ and $d_y=-\alpha_hk^3\cos(3\phi)$,
and the Berry phase is found to be $3\pi$. However, only at low
density approximation, the spin torque-Hall conductivity may be
numerically determined by the Berry phase~\cite{k3hole}. For the
$k$-cubic Rashba-Dresselhaus system~\cite{Bulaev05}, we have
$d_x=[\alpha_h\sin(3\phi)+\beta_h\cos\phi]k^3$ and
$d_y=[-\alpha_h\sin(3\phi)+\beta_h\sin\phi]k^3$. The corresponding
Berry phase is $3\pi$ for $\alpha_h^2>\beta_h^2$ and $\pi$ for
$\alpha_h^2<\beta_h^2$. It has been found that the spin
torque-Hall conductivity at low density approximation is
$-9e/4\pi$ for $\alpha_h^2>\beta_h^2$ and $-3e/4\pi$ for
$\alpha_h^2<\beta_h^2$~\cite{k3hole2}, which can be numerically
determined by the Berry phase. That is, the relationship between
the spin torque current and the Berry phase would depend on the
position of the Fermi level when the higher momentum is included
in the system. Interestingly, it has recently been shown that the
Fermi surface contribution of the pure spin torque response Eq.
(\ref{Sigmatau}) in two-dimensional systems including higher
momentum always exhibits quantized conductivity~\cite{Raichev}.

On the other hand, it seems that a non-vanishing Berry phase on
the Fermi surface implies a non-vanishing charge Hall conductance
in a time-reversal symmetric system. We can choose a new basis
vector such that the charge Hall conductance
vanishes~\cite{MCChang05}, however, the physics conclusion Eq.
(\ref{TorSpinMatrix}) [see also Eq. (\ref{TorSpin-linear})] is
still unchanged. The only change is that the Berry phase in the
new basis would depend on the band index, which could be
experimentally determined~\cite{Murakawa2013}.

%We now discuss the experiment on detecting the spin torque current
%in the spin-Hall effect. The spin torque current stems from the
%variation of spin of local electron. As a result, the spin torque
%current does not accompany the charge flow. In this sense, pure
%electric field without current injection could generate only
%torque-spin current. When an electron cannot be accelerated, the
%spin $z$ component cannot be tilted up by applying an electric
%field except at $\mk=0$. As shown in this paper, remarkably, the
%Berry curvature is a Dirac delta function peaked at $\mk=0$ that
%largely contributes to the torque-spin current. This is
%schematically depicted in Fig. \ref{FigExp}. The experiment we
%proposed can also be used to the system where the torque-spin
%current is not purely caused by the Berry curvature. For example,
%it can be shown that the Berry curvature in the k-cubic Rashba
%coupling~\cite{Winkler02} and k-cubic Rahshba-Dresselhaus
%coupling~\cite{Bulaev05} is also Dirac-delta functions peaked at
%$\mk=0$.

\section{Conclusion}\label{sec:Conclusion}
For a generic $k$-linear spin-orbit-coupled system, we calculated
the Berry phase on the Fermi surface, and the conventional and
spin torque conductivities by using the Kubo formula. The
conventional spin Hall current in general depends on the
spin-orbit strength and is not proportional to the Berry phase.
The longitudinal term of the conventional spin conductivity also
does not vanish.

We found that the spin torque Hall current is proportional to the
Berry phase and that the longitudinal spin torque current
vanishes. Since the Berry phase effect prohibits the longitudinal
response and results in  antisymmetric properties in transverse
conductivities, in this sense, we found that the spin torque
conductivity is truly caused by the Berry phase effect on the
Fermi surface. We showed that the Berry phase on the Fermi surface
manifests two invariant quantities. The Berry phase is a sign
function of the spin-orbit matrix, which is invariant under
rotation. This means that the magnitude of the spin torque current
response is independent of the direction of the applied electric
field. The Berry phase also implies that a small fluctuation of
spin-orbit strength (if it does not cause the phase transition)
does not change the magnitude of the spin torque current. The
Berry phase divided by $\pi$ is an integer, $+1$ or $-1$, which is
a topological number. The topological number $\pm1$ due to the
Berry phase not only represents the different phase of the spin
Hall system but also reflects the rotationally invariant property.
The change in the topological number results in the phase
transition of exchanging the orbital chirality of spin $s_z$ and
$-s_z$ in the spin Hall system. The phase transition phenomenon
occurs in conventional spin and spin torque currents.

Because the Fermi level does not lie in the true gap, the
topological number would be destroyed by the influence of impurity
scattering, which depends on the scattering mechanism.

Hopefully, our interesting predictions of the invariant properties
of the spin torque current will stimulate measurements in
two-dimensional semiconductor systems in the near future.

\section*{ACKNOWLEDGMENTS}
T. W. C. would like to thank Guang-Yu Guo for valuable discussions
on the spin current. The authors thank Ministry of Science and
Technology for the financial support under Grant No. NSC
101-2112-M-110-013-MY3.

\appendix
\section{Torque-Spin conductivity}\label{App-TorSpin}
In this appendix, we derive Eq. (\ref{TorSpin}). Since in
obtaining $\sigma^{\tau_z}_{ab}$, we have to take the limit
$q_a\rightarrow0$, we expand each term to first order of $q_a$. By
calculating the matrix element $\langle
n\mk|\tau_z(\mk,\mq)|-n\mk+\mq\rangle$, where we have replaced
$n'$ by $-n$,
\begin{equation}\label{App-tau}
\begin{split}
&\langle
n\mk|\tau_z(\mk,\mq)|-n\mk+\mq\rangle\\
&=-in\left(d+\frac{q_a}{2}\frac{\partial d}{\partial
k_a}\right)+q_a\frac{nd}{2}\frac{\partial\theta}{\partial
k_a}+o(q_a^2).
\end{split}
\end{equation}
For the matrix element $\langle
-n\mk|v_b(\mk,\mq)|n\mk+\mq\rangle$, we have
\begin{equation}\label{App-v}
\begin{split}
&\langle -n\mk+\mq|v_b(\mk,\mq)|n\mk\rangle\\
&=\frac{ind}{\hbar}\frac{\partial\theta}{\partial
k_b}+\frac{-in}{2}q_a\left(\frac{d_y}{d}\frac{\partial^2
d_x}{\hbar\partial k_a\partial k_b}+\frac{d_x}{d}\frac{\partial^2
d_y}{\hbar\partial k_a\partial k_b}\right)\\
&+q_a\frac{\partial\theta}{\partial
k_a}\left(\frac{i}{2}\frac{\partial
\varepsilon_{\mk}}{\hbar\partial k_a}-\frac{n}{2}\frac{\partial
d_x}{\hbar\partial k_b}e^{i\theta}+\frac{in}{2}\frac{\partial
d_y}{\hbar\partial k_b}e^{i\theta}\right)+o(q_a^2).
\end{split}
\end{equation}
On the other hand, we have
\begin{equation}\label{App-f}
\begin{split}
&\frac{f_{n\mk}-f_{-n\mk+\mq}}{(\mathE_{n\mk}-\mathE_{-n\mk+\mq})^2}\\
&=\frac{f_{n\mk}-f_{-n\mk}}{(\mathE_{n\mk}-\mathE_{-n\mk})^2}\\
&+q_a\frac{\partial\mathE_{-n\mk}}{\partial
k_a}\left[2\frac{f_{n\mk}-f_{-n\mk}}{(\mathE_{n\mk}-\mathE_{-n\mk})^3}-\frac{\partial
f_{-n\mk}/\partial\mathE_{-n\mk}}{(\mathE_{n\mk}-\mathE_{-n\mk})^2}\right]+o(q_a^2).
\end{split}
\end{equation}
Inserting Eqs. (\ref{App-tau}), (\ref{App-v}) and (\ref{App-f})
into Eq. (\ref{Kubo-Torspin-1}) and after straightforward
calculations, we have
\begin{equation}\label{App-sigma}
\begin{split}
\sigma^{s_z}_{ab}&=-\frac{e}{V}\sum_{\mk}\frac{1}{4}\frac{\partial\theta}{\partial
k_b}\sum_{n}\left(\frac{\partial
f_{\mk+}}{\partial\mathE_{\mk+}}\frac{\partial\mathE_{+\mk}}{\partial
k_a}+\frac{\partial
f_{\mk-}}{\partial\mathE_{\mk-}}\frac{\partial\mathE_{-\mk}}{\partial
k_a}\right)\\
&+\frac{e}{V}\sum_{\mk}\frac{f_{\mk+}-f_{\mk-}}{4d}\frac{\partial\varepsilon_{\mk}}{\partial
k_b}\frac{\partial\theta}{\partial k_a}\\
&-\frac{e}{V}\sum_{\mk}\frac{f_{\mk+}-f_{\mk-}}{2d}\frac{\partial\varepsilon_{\mk}}{\partial
k_a}\frac{\partial\theta}{\partial k_b}.
\end{split}
\end{equation}
The first term of Eq. (\ref{App-sigma}) is the pure spin torque
conductivity $\Sigmatau_{ab}$ [see Eq. (\ref{Sigmatau})]. The
second and third terms of Eq. (\ref{App-sigma}) equals to
$\sigma^{s_z}_{ba}$, and $-2\sigma^{s_z}_{ab}$, respectively.

\section{Calculation of $\Sigmatau_{ab}$}\label{App-Sigmatau}

In this appendix, we want to calculate the response
$\Sigmatau_{ab}$ [Eq. (\ref{Sigmatau})] in the system with generic
$k$-linear Hamiltonian. At zero temperature, the Fermi level $\mu$
determines the Fermi momentum $k_{F}^{\pm}$ for the two bands by
$\mu=\frac{\hbar^2(k_F^n)^2}{2m}-nk_F^n\Gamma$. We have
\begin{equation}\label{AppC-Fermik}
k_F^{\pm}=\pm\frac{m\Gamma}{\hbar^2}+\frac{m}{\hbar^2}\sqrt{\Gamma^2+\frac{2\mu\hbar^2}{m}}.
\end{equation}
The term $\partial f_{n\mk}/\partial\mathE_{n\mk}$ is given by the
delta function $-\delta(\mu-\mathE_{n\mk})$. The term
$\mu-\mathE_{n\mk}$ can be written as
$-\frac{\hbar^2}{2m}(k-k_F^n)(k+k_F^n-\frac{2nm\Gamma}{\hbar^2})$.
Consider $n=+1$, we have
$(k_F^+-2m\Gamma/\hbar^2)=(m\sqrt{\Gamma^2+2\mu\hbar^2/m}/\hbar^2-m\Gamma/\hbar^2)=k_F^-$.
On the other hand, when $n=-1$, we have
$(k_F^--2m\Gamma/\hbar^2)=k_F^+$. Using
$\delta(x-a)(x-b)=[\delta(x-a)-\delta(x-b)]/|a-b|$ and Eq.
(\ref{AppC-Fermik}), we have
\begin{equation}\label{delta1}
\begin{split}
\frac{\partial
f_{n\mk}}{\partial\mathE_{n\mk}}&=-\delta(\mu-\mathE_{n\mk})\\
&=-\frac{1}{\sqrt{\Gamma^2+\frac{2\mu\hbar^2}{m}}}\left[\delta(k-k_F^n)-\delta(k+k_F^{-n})\right].
\end{split}
\end{equation}
Since $k>0$, the second term of Eq. (\ref{delta1}) vanishes, and
thus, we have
\begin{equation}\label{delta2}
\frac{\partial
f_{n\mk}}{\partial\mathE_{n\mk}}=-\frac{\delta(k-k_F^n)}{\sqrt{\Gamma^2+\frac{2\mu\hbar^2}{m}}}=-\frac{2m}{\hbar^2}\frac{\delta(k-k_F^n)}{k_F^++k_F^-}.
\end{equation}
Using Eqs. (\ref{delta2}), (\ref{Sigmatau}) and
(\ref{BerryA-linear}), the term $\Sigmatau_{ab}$ is given by
\begin{equation}\label{AppC-Sigmatau1}
\begin{split}
\Sigmatau_{ab}&=-\frac{e}{2V}\sum_{n\mk}\frac{\partial
f_{n\mk}}{\partial k_a}A_b\\
&=-\frac{e}{2V}\sum_{\mk}\left(\frac{\partial
f_{\mk+}}{\partial\mathE_{\mk+}}\frac{\partial\mathE_{\mk+}}{\partial
k_a}+\frac{\partial
f_{\mk-}}{\partial\mathE_{\mk-}}\frac{\partial\mathE_{\mk-}}{\partial
k_a}\right)A_b\\
&=\frac{e}{8\pi^2}\int_0^{2\pi}d\phi\frac{2m/\hbar^2}{k_F^++k_F^-}\left\{f_1(\phi)-f_2(\phi)\right\},
\end{split}
\end{equation}
where
\begin{equation}
\begin{split}
f_1(\phi)&=\int kdk\frac{\partial\varepsilon_{\mk}}{\partial
k_a}A_b\left[\delta(k-k_F^-)+\delta(k-k_F^+)\right],\\
f_2(\phi)&=\int kdk\frac{\partial d}{\partial
k_a}A_b\left[\delta(k-k_F^-)-\delta(k-k_F^+)\right].
\end{split}
\end{equation}
It can be straightforwardly shown that $f_2(\phi)$ always vanishes
for $a=x,y$ and $b=x,y$. This can also be seen as follows. Because
we use the $\mk$-linear system, we have $d\sim k$ and
$A_b\sim1/k$, and thus, the term $k(\partial d/\partial k_a)A_b$
is a function of $\phi$ only. Therefore, for $\Sigmatau_{ab}$, we
have
\begin{equation}\label{AppC-Sigmatau2}
\begin{split}
\Sigmatau_{ab}=&\frac{e}{8\pi^2}\int_0^{2\pi}d\phi\frac{2}{k_F^++k_F^-}\int
kdkk_aA_b\\
&\times\left[\delta(k-k_F^-)+\delta(k-k_F^+)\right].
\end{split}
\end{equation}
Using the Berry vector potential for $k$-linear system [see Eq.
(\ref{BerryA-linear})], the matrix formed by $k_aA_b$ is given by
\begin{equation}\label{AppC-kA}
\left(\begin{array}{cc}
k_xA_x&k_xA_y\\
k_yA_x&k_yA_y
\end{array}\right)=\frac{\detbeta}{2\Gamma^2}\left(\begin{array}{cc}
-\sin\phi\cos\phi&\cos^2\phi\\
-\sin^2\phi&\sin\phi\cos\phi
\end{array}\right).
\end{equation}
Inserting  Eq. (\ref{AppC-kA}) into Eq. (\ref{AppC-Sigmatau2}), we
have
\begin{equation}\label{App-Sigmatau}
\begin{split}
&\left(\begin{array}{cc}
\Sigmatau_{xx}&\Sigmatau_{xy}\\
\Sigmatau_{yx}&\Sigmatau_{yy}
\end{array}\right)\\
&=\frac{e}{8\pi^2}\detbeta\int_0^{2\pi}d\phi\frac{1}{\Gamma^2}\left(\begin{array}{cc}
-\sin\phi\cos\phi&\cos^2\phi\\
-\sin^2\phi&\sin\phi\cos\phi
\end{array}\right).
\end{split}
\end{equation}
Compare Eq. (\ref{App-Sigmatau}) with Eq. (\ref{ConSpin-linear}),
we find that $\Sigmatau_{ab}$ is equal to $\sigma^{s_z}_{ab}$ for
$a=x,y$ and $b=x,y$.


\begin{thebibliography}{99}
\bibitem{para} The Fermi level is 2.67 meV and the partilce mass is 0.08 in
units of the bare electron mass. For the energy dispersion in Fig.
1(a), we use $\beta_{xx}=4.69\times10^{-2}$eV nm,
$\beta_{xy}=3.88\times10^{-2}$eV nm,
$\beta_{yx}=3.21\times10^{-2}$ eV nm,
$\beta_{yy}=5.59\times10^{-2}$ eV nm. For the energy dispersion in
Fig. 1(b), we use $\beta_{xx}=3\times10^{-2}$eV nm,
$\beta_{xy}=6\times10^{-2}$eV nm, $\beta_{yx}=2\times10^{-2}$ eV
nm, $\beta_{yy}=4\times10^{-2}$ eV nm.
\end{thebibliography}
\end{document}